\documentclass[pra,aps,showpacs,twocolumn]{revtex4}

\usepackage{amsmath}
\usepackage{bm}
\usepackage{graphicx}

\begin{document}

\title{Stabilization of BEC droplet in free space by feedback control of
interatomic interaction}

\author{Hiroki Saito$^1$}
\author{Masahito Ueda$^{2,3}$}
\affiliation{$^1$Department of Applied Physics and Chemistry, The
University of Electro-Communications, Tokyo 182-8585, Japan \\
$^2$Department of Physics, Tokyo Institute of Technology,
Tokyo 152-8551, Japan \\
$^3$ERATO, Japan Science and Technology Corporation (JST), Saitama
332-0012, Japan
}

\date{\today}

\begin{abstract}
A self-trapped Bose-Einstein condensate in three-dimensional free space is
shown to be stabilized by feedback control of the interatomic interaction
through nondestructive measurement of the condensate's peak column
density.
The stability is found to be robust against poor resolution and
experimental errors in the measurement.
\end{abstract}


\maketitle

\section{Introduction}

Matter--wave bright solitons of a Bose-Einstein condensate (BEC) in a
quasi-one dimensional (1D) trapping potential have been realized by the
ENS group~\cite{Khay} and the Rice group~\cite{Strecker}.
The stability of this self-trapped state is achieved by the balance
between the attractive interatomic interaction and the quantum kinetic
pressure.
In 1D, the bright soliton is stable and robust against noise.
In 2D and 3D, however, the balance between the attractive interaction and
the kinetic pressure is precarious, and infinitesimal deviations from the
stationary state cause a collapse or expansion of the system.
While self-trapped liquids, such as a raindrop, are quite common, a
self-trapped gas is a novel state of matter.
Such a stable self-trapped state in a gaseous BEC in 2D or 3D, if it can
be realized, might be referred to as a matter-wave droplet or simply a BEC
droplet.

A scheme proposed in Refs.~\cite{SaitoL, Ab03} to stabilize a BEC droplet
is to oscillate the interatomic interaction rapidly using Feshbach
resonance~\cite{Inouye,Cornish}.
The rapid oscillation of the interaction produces an effective potential
that prevents the condensate from collapsing.
This phenomenon is similar to the stabilization of an inverted pendulum
by an oscillating pivot~\cite{Landau}.
Several researchers have demonstrated the stabilization of a BEC droplet
in 2D with oscillating interactions by numerically solving the
Gross--Pitaevskii (GP) equation~\cite{SaitoL, Ab03, Montesinos, Itin}.
Montesinos {\it et al.}~\cite{GasparL} have shown that this stabilization
in 2D is also possible for a multicomponent BEC.
Matuszewski {\it et al.}~\cite{Trippen} have predicted 3D breather
solitons confined in a 1D lattice.
However, in 3D free space, it appears that an oscillating interaction
alone cannot stabilize a BEC droplet due to dynamical
instabilities~\cite{Ab03,Montesinos}.
By taking into account the effect of energy dissipation, which always
exists in realistic situations, we have shown that a BEC droplet with
oscillating interactions can be stabilized in 3D~\cite{SaitoA}.

In the present paper, we show that a BEC droplet in 3D free space can be
stabilized by feedback control of the interaction through nondestructive
measurement of the condensate's peak column density.
Real-time monitoring of the density profile of a condensate is possible by
using a nondestructive in-situ imaging method~\cite{Andrews}.
The collapse or expansion of the condensate can be prevented by a decrease
or increase, respectively, in the strength of the attractive interaction
if the peak density of the condensate increases above or decreases below a
prescribed value.
Thus, the shape of the BEC droplet can be maintained without collapse or
expansion by negative feedback from the result of the real-time
measurement.
We will also examine the effect of experimental imperfections in the
real-time measurement, such as spatial and time resolutions and
experimental errors, and show that the stabilized BEC droplet is robust
against these imperfections.

This paper is organized as follows.
Section~\ref{s:GP} discusses the stationary solutions of the GP equations
in 1D, 2D, and 3D free space.
Section~\ref{s:result} numerically investigates the dynamics of the
condensate under feedback control, and shows that a BEC droplet in 3D can
be stabilized for a wide range of parameters.
Section ~\ref{s:var} presents variational analysis using a Gaussian
trial function.
Section~\ref{s:noise} studies the stability against measurement errors and 
finite measurement resolution, and shows that a BEC droplet is
robust against these experimental imperfections.
Finally, Sec.~\ref{s:conc} concludes this paper.

\section{Stationary solutions of the Gross--Pitaevskii equation in free
space}
\label{s:GP}

We first consider a stationary state of a BEC with an attractive
interaction in free space.
The dynamics of the BEC are described by the GP equation,
\begin{equation} \label{GP}
i \hbar \frac{\partial \psi}{\partial t} = -\frac{\hbar^2}{2m} \nabla^2
\psi + V \psi + \frac{4\pi \hbar^2 a}{m} |\psi|^2 \psi,
\end{equation}
where $m$ is the mass of an atom, $V$ is the external potential, and $a$
is the $s$-wave scattering length.
The wave function is normalized with $\int |\psi|^2 d{\bm r} = N$, with
$N$ being the number of atoms.

When the external potential is given by $V = m \omega_{\rm 1d}^2 (x^2 +
y^2) / 2$ and $\hbar \omega_{\rm 1d}$ is much larger than the other
characteristic energies, the degrees of freedom of the BEC in the $x$ and
$y$ directions are frozen and the system behaves as an effective 1D
system.
Writing the wave function as
\begin{equation}
\psi(\bm{r}) = \sqrt{\frac{m \omega_{\rm 1d}}{\pi \hbar}} e^{-\frac{m
\omega_{\rm 1d}}{2 \hbar} (x^2 + y^2) - i \omega_{\rm 1d} t} \psi_z(z),
\end{equation}
and integrating Eq.~(\ref{GP}) over $x$ and $y$,
we obtain an effective 1D equation:
\begin{equation}
i \hbar \frac{\partial \psi_z}{\partial t} = -\frac{\hbar^2}{2m}
\frac{\partial^2 \psi_z}{\partial z^2} + g_{\rm 1d} |\psi_z|^2 \psi_z,
\end{equation}
where $g_{\rm 1d} = 2 \hbar \omega_{\rm 1d} a$.
If $a < 0$, this equation has a well-known soliton
solution~\cite{Malomed}:
\begin{equation}
\psi_z = e^{-i \mu t / \hbar} \sqrt{\frac{\eta N}{2}}\ {\rm sech} \eta (z
- z_0),
\end{equation}
where $\mu = -m \omega_{\rm 1d}^2 a^2 N^2 / 2$, $\eta = m \omega_{\rm 1d}
|a| N / \hbar$, and $z_0$ is the position of the peak.
This is the ground state solution, and it has no dynamical instabilities.
Thus an attractive BEC is stable in 1D.

When $V = m \omega_{\rm 2d}^2 z^2 / 2$ and $\hbar \omega_{\rm 2d}$ is much
larger than the other characteristic energies, the system behaves as an
effective 2D system.
Substituting the wave function
\begin{equation}
\psi(\bm{r}) = \left( \frac{m \omega_{\rm 2d}}{\pi \hbar} \right)^{1/4}
e^{-\frac{m \omega_{\rm 2d}}{2 \hbar} z^2 - i \frac{\omega_{\rm 2d}}{2} t}
\psi_{xy}(x, y)
\end{equation}
into Eq.~(\ref{GP}), and integrating the result over $z$, we obtain an
effective 2D equation:
\begin{equation} \label{GP2D}
i \hbar \frac{\partial \psi_{xy}}{\partial t} = -\frac{\hbar^2}{2m} \left(
\frac{\partial^2}{\partial x^2} + \frac{\partial^2}{\partial y^2} \right)
\psi_{xy} + g_{\rm 2d} |\psi_{xy}|^2 \psi_{xy},
\end{equation}
where $g_{\rm 2d} = (8 \pi \hbar^3 a^2 \omega_{\rm 2d} / m)^{1/2}$.
At the critical strength of the interaction
\begin{equation} \label{g2dcr}
g_{\rm 2d}^{\rm cr} \simeq -1.862 \frac{\pi \hbar^2}{m N},
\end{equation}
Eq.~(\ref{GP2D}) has a stationary self-trapped solution, which is known as
the Townes soliton~\cite{Townes}.
If the peak of the Townes soliton is located at the origin ${\bm r} = 0$,
the wave function is axisymmetric and the density $|\psi|^2$ monotonically
decreases to zero for $r \rightarrow \infty$.
The Townes soliton also has a scaling property;
if $\psi_{xy}(x, y)$ is a stationary solution of Eq.~(\ref{GP2D}), the
scaled wave function $\alpha \psi_{xy}(\alpha x, \alpha y)$ with an
arbitrary scaling parameter $\alpha$ is also a stationary solution of
Eq.~(\ref{GP2D}).
However, the Townes soliton is dynamically unstable in the sense that an
infinitesimal deviation from the stationary solution grows exponentially
in time, and therefore the Townes soliton eventually collapses or
expands.

When $V = 0$, i.e., in 3D free space, Eq.~(\ref{GP}) with $a < 0$ has
a stationary self-trapped solution that is dynamically unstable like the
Townes soliton.
The density of this stationary solution has spherical symmetry and
monotonically decreases to zero for $r \rightarrow \infty$~\cite{Sulem}.
The striking difference between this 3D stationary state and the Townes
soliton is in their scaling properties.
Normalizing the length, time, and wave function in units of $\ell$, $m
\ell^2 / \hbar$, and $(N / \ell^3)^{1/2}$, respectively, where $\ell$ is
an arbitrary length scale, Eq.~(\ref{GP}) reduces to
\begin{equation} \label{GPn}
i \frac{\partial \psi}{\partial t} = -\frac{\nabla^2}{2} \psi + g |\psi|^2
\psi,
\end{equation}
where $g = 4\pi N a / \ell$.
Therefore, if $\psi({\bm r}, t)$ is a solution of the GP equation with
scattering length $a$, the scaled wave function $\alpha^{3/2} \psi(\alpha
{\bm r}, \alpha^2 t)$ also satisfies the GP equation with scattering
length $a / \alpha$.
This indicates that there always exists an unstable stationary solution
for any $g < 0$, and the solutions for different $g$'s are related by the
scaling transformation.
In contrast, in 2D, the Townes soliton exists only for a particular value
of $g_{\rm 2d} = g_{\rm 2d}^{\rm cr}$ (see Eq.~(\ref{g2dcr})) for an
arbitrary scaling parameter $\alpha$.

\begin{figure}[tb]
\includegraphics[width=8.4cm]{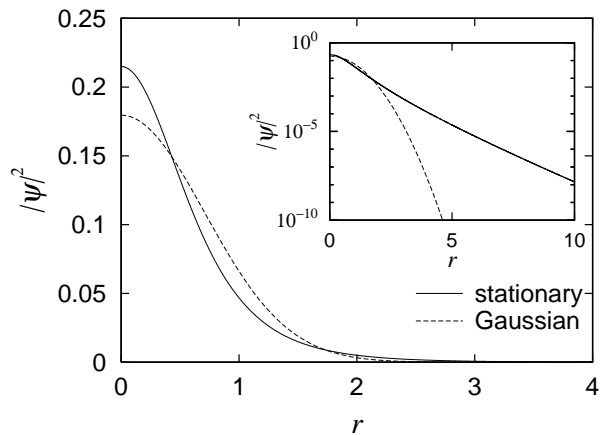}
\caption{
Stationary unstable solution of Eq.~(\ref{GPn}) (solid curve) for $g
= -(2 \pi)^{3/2}$.
The Gaussian function $e^{-r^2} / \pi^{3/2}$ is superimposed as a dashed
curve for comparison.
See Sec.~\ref{s:var} for the choice of parameters.
The inset shows the same functions on a logarithmic scale to show the
difference in the large-$r$ behavior.
}
\label{f:townes}
\end{figure}

In Fig.~\ref{f:townes}, we plot the density profile of the unstable
stationary state in 3D for $g = -(2 \pi)^{3/2}$, which is numerically
obtained by the Newton--Raphson method~\cite{Edwards}.
The inset of Fig.~\ref{f:townes} shows the logarithmic plot of the
large-$r$ behavior.
We find that the tail of the unstable stationary state (solid curve) is
longer than that of the Gaussian wave function $e^{-r^2} / \pi^{3/2}$
(dashed curve).
Generally, in $d$ dimensions the unstable stationary wave function has an
asymptotic form $r^{(1 - d) / 2} e^{-c r}$ with constant $c$ for $r
\rightarrow \infty$~\cite{Sulem}.
The dependence of the tail on $r$ in the inset of Fig.~\ref{f:townes} is
consistent with this functional form with $d = 3$ and $c \simeq 1.3$.

\section{Feedback control of a BEC droplet}
\label{s:result}

The 3D stationary solution of Eq.~(\ref{GPn}) is dynamically unstable
against collapse and expansion.
The aim of the present paper is to show that we can stabilize it by
controlling the scattering length $a$.
When the system is about to collapse, we can increase $a$ to prevent the
collapse.
Similarly, a decrease in $a$ can prevent expansion.
Thus, by measuring the density profile of the condensate in a
nondestructive manner, we can achieve feedback control of $a$ to stabilize
a BEC droplet.

An observable quantity in nondestructive phase-contrast
imaging~\cite{Andrews} is the column density of the condensate, given by
\begin{equation} \label{dc}
d_{\rm col}(x, y, t) = \int dz |\psi(\bm{r}, t)|^2,
\end{equation}
where the line of sight is assumed to be in the $z$ direction.
We use the peak value of the column density,
\begin{equation} \label{D}
D \equiv d_{\rm col}(0, 0, t),
\end{equation}
for the feedback control.

Let $D_0$ be the target value of the stabilized column density.
The feedback loop should operate on the strength of the interaction $g$ in
such a manner that any deviation $\tilde D \equiv D - D_0$ from the target
value $D_0$ will be suppressed.
We assume that the time derivative of $g$ depends on $\tilde D$ up to the
second derivative with respect to time:
\begin{equation} \label{feedback}
\dot{g} = A (D - D_0) + B \dot{D} + C \ddot{D},
\end{equation}
where $A$, $B$, and $C$ are dimensionless constants.
Recovering the dimensions of $D$ and $t$ by multiplying by $N / \ell^2$ and
$m \ell^2 / \hbar$, respectively, where $\ell$ is an arbitrary unit of
length, we can write Eq.~(\ref{feedback}) as
\begin{equation}
\dot{a} = \frac{\ell \hbar}{4 \pi N^2 m} A (D - D_0) + \frac{\ell^3}{4 \pi
N^2} B \dot{D} + \frac{m \ell^5}{4 \pi N^2 \hbar} C \ddot{D}.
\end{equation}

In experiments, a time sequence of phase-contrast images is taken by a CCD
camera at a certain frame rate (e.g., 20 kHz in Ref.~\cite{Higbie}).
Therefore, the measurement of $D$ is performed at discrete times and the
time derivatives in Eq.~(\ref{feedback}) are approximately obtained by
\begin{subequations} \label{discrete}
\begin{eqnarray}
\label{discrete1}
\dot{D}(t) & \simeq & \frac{1}{\Delta t} [D(t) - D(t - \Delta t)], \\
\label{discrete2}
\ddot{D}(t) & \simeq & \frac{1}{\Delta t^2} [D(t) - 2 D(t - \Delta t) +
D(t - 2 \Delta t)], \nonumber \\
\end{eqnarray}
\end{subequations}
where $\Delta t$ is the interval between measurements.
Accordingly, $g$ is also changed stepwise as
\begin{equation} \label{gdiscrete}
g(t) = g(t - \Delta t) + \Delta t \left\{ A [D(t) - D_0] + B \dot{D}(t) +
C \ddot{D} (t) \right\},
\end{equation}
where Eqs.~(\ref{discrete1}) and (\ref{discrete2}) are used for
$\dot{D}(t)$ and $\ddot{D}(t)$.
The interval $\Delta t$ must be made much smaller than the characteristic
time scale of the dynamics.
The $\Delta t$ dependence of the stability is discussed in
Sec.~\ref{s:noise}.

We assume that the initial state is the noninteracting ground state in an
isotropic harmonic potential $V = m \omega^2 \bm{r}^2 / 2$.
Henceforth, we take the units of length and time to be $\ell = [\hbar / (m
\omega)]^{1/2}$ and $\omega^{-1}$, respectively.
The initial state is then described by the Gaussian wave function $\psi =
e^{-r^2/2} / \pi^{3/4}$.
At $t = 0$, the trapping potential is suddenly switched off and the
strength of the interaction is set to be $g = -(2 \pi)^{3/2}$.
The strength of the interaction $g$ evolves in time according to
Eq.~(\ref{gdiscrete}), where the value of $D_0$ is chosen to be the peak
column density of the initial wave function, $D_0 = \int dz |\psi(x = 0, y
= 0, z, t = 0)|^2 = 1 / \pi$.
The wave function evolves in time according to Eq.~(\ref{GPn}), which is
numerically solved by the Crank--Nicolson method.
We neglect the effect of gravity by assuming that it is canceled using,
e.g., a technique of magnetic levitation~\cite{Herbig}.

\begin{figure}[tb]
\includegraphics[width=8.4cm]{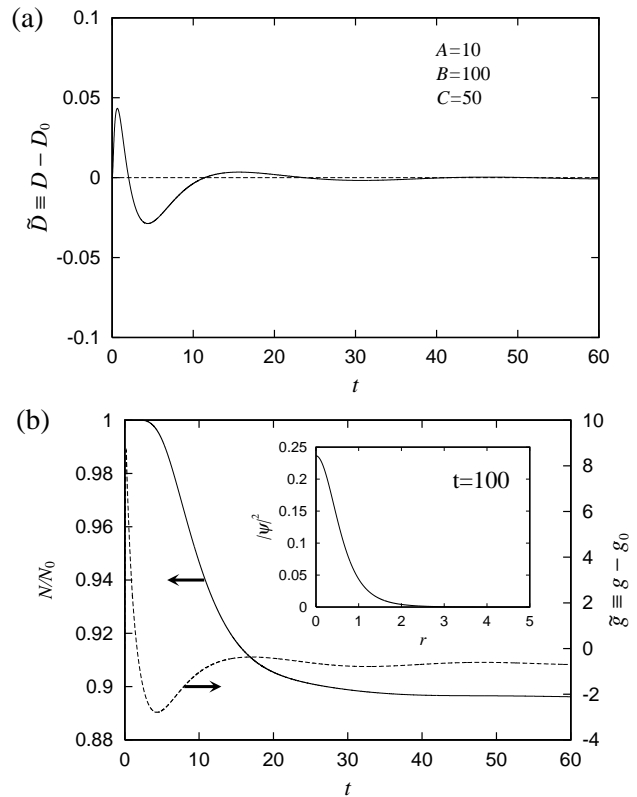}
\caption{
Time evolution of (a) the deviation $\tilde D \equiv D - D_0$ of the peak
column density (see Eqs.~(\ref{dc}) and (\ref{D})) and (b) the fraction of
remaining atoms $N / N_0$ (solid curve) and the strength of the
interaction $\tilde g = g - g_0$ (dashed curve), for $A = 10$, $B = 100$,
$C = 50$, $\Delta t = 0.01$, $D_0 = 1 / \pi$, and $g_0 = -(2 \pi)^{3/2}$.
The initial state is the Gaussian wave function $\psi = e^{-r^2/2} /
\pi^{3/4}$, and the value of $\tilde g$ starts from $0$ (not visible in
the present resolution).
The inset in (b) shows the converged density profile $|\psi|^2$ at $t =
100$.
}
\label{f:evolution}
\end{figure}

Figure~\ref{f:evolution} (a) shows the time evolution of $D$ for $A = 10$,
$B = 100$, $C = 50$, and $\Delta t = 0.01$.
The value of $D$ deviates significantly from $D_0$ only for $t \lesssim
20$, and quickly converges to $D_0$ thereafter.
This indicates that the feedback control of the interaction successfully
works to stabilize a BEC droplet in 3D.
The inset in Fig.~\ref{f:evolution} (b) shows the density profile of the
condensate at $t = 100$.
This converged density profile is found to be related to the stationary
solution of Eq.~(\ref{GPn}) by appropriate scaling.
The feedback control can thus transform a Gaussian wave function into a
stationary solution of Eq.~(\ref{GPn}).
During this transformation, a certain fraction of atoms are lost from the
condensate.
The solid curve in Fig.~\ref{f:evolution} (b) shows the fraction of atoms
remaining within the radius $r = 10$ around the center of the condensate,
\begin{equation}
\frac{N}{N_0} = \int_0^{10} dr 4 \pi r^2 |\psi|^2.
\end{equation}
We find that roughly 10\% of atoms are scattered away from the condensate
without returning to the center due to the lack of a trapping potential.

\begin{figure}[tb]
\includegraphics[width=8.4cm]{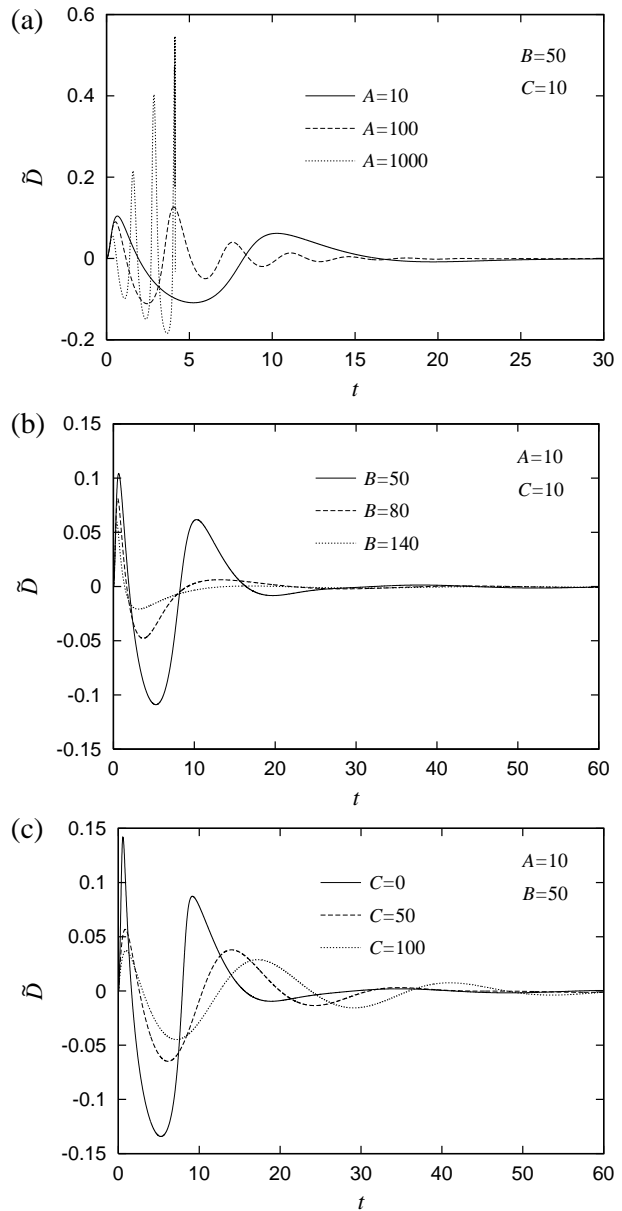}
\caption{
Dependence of the dynamics of $D$ on the feedback parameters $A$, $B$, and
$C$.
The other conditions are the same as in Fig.~\ref{f:evolution}.
}
\label{f:depend}
\end{figure}

Figure~\ref{f:depend} shows the dependence of the dynamics on the feedback
parameters.
Figure~\ref{f:depend} (a) shows that the parameter $A$ controls the
amplitude of the density oscillation.
The oscillation frequency increases with an increase in $A$, and for
large $A$ the system becomes unstable against the growth in the amplitude
of the density oscillations.
This is because the term of $A$ in Eq.~(\ref{feedback}) plays a role of
pulling the value of $D$ back to $D_0$.
If $A$ is too large, this pulling causes an overshoot, resulting in
oscillations of $D$.
Figures~\ref{f:depend} (b) and (c) indicate that the amplitudes of the
oscillations decrease with increasing $B$ and $C$.
The decay of the oscillations accelerates for larger $B$
[Fig.~\ref{f:depend} (b)], while it decelerates for larger $C$
[Fig.~\ref{f:depend} (c)].

\begin{figure}[tb]
\includegraphics[width=8.4cm]{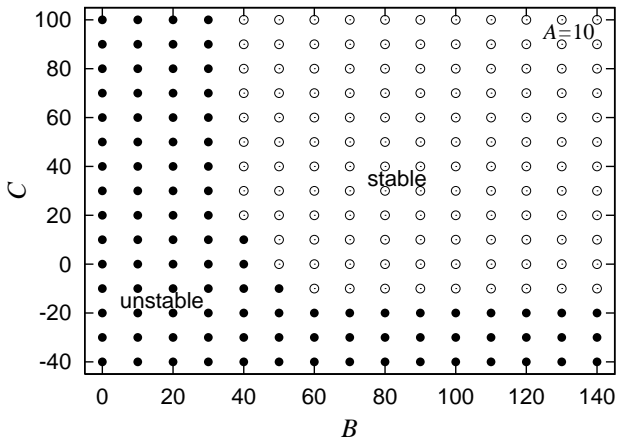}
\caption{
Stability diagram of the BEC droplet under feedback control for $A = 10$.
The other conditions are the same as in Fig.~\ref{f:evolution}.
The open circles show stable points and the filled circles show unstable
points.
}
\label{f:stability}
\end{figure}

Figure~\ref{f:stability} shows a stability diagram of the feedback control
with respect to the feedback parameters $B$ and $C$.
We take the value $A = 10$, since larger values of $A$ cause rapid
oscillations of the system, as shown in Fig.~\ref{f:depend} (a).
We find that the BEC droplet can be stabilized for $B \gtrsim 30$ and $C
\gtrsim -20$.

\section{Variational analysis}
\label{s:var}

In order to understand in an analytic manner the stability of the system
subject to the feedback control discussed in Sec.~\ref{s:result}, we
conduct a variational analysis.

We employ the Gaussian trial function~\cite{Garcia}
\begin{equation} \label{Gauss}
\psi_{\rm G} = \frac{1}{\pi^{3/4} R^{3/2}} e^{-\frac{r^2}{2 R^2} + i
\frac{\dot{R}}{2 R} r^2},
\end{equation}
where $R(t)$ is a variational parameter that characterizes the size of the
condensate.
Equation~(\ref{GPn}) is derived by the application of the variational
principle to the action
\begin{equation} \label{S}
S = \int d \bm{r} dt \psi^* \left( i \frac{\partial}{\partial t} +
\frac{\nabla^2}{2} - \frac{g}{2} |\psi|^2 \right) \psi.
\end{equation}

Substituting the variational wave function (\ref{Gauss}) into
Eq.~(\ref{S}) and minimizing the result with respect to $R$, we obtain the
equation of motion for $R$ as
\begin{equation} \label{Rddot}
\ddot{R} = -\frac{1}{R^3} - \frac{g}{(2 \pi)^{3/2}} \frac{1}{R^4}.
\end{equation}
The unstable stationary solution of Eq.~(\ref{Rddot}) is obtained for $R =
-g / (2 \pi)^{3/2}$; hence, if $g = -(2 \pi)^{3/2}$, we have $R = 1$.
In Fig.~\ref{f:townes}, we plot the Gaussian function (\ref{Gauss}) with
$R = 1$ as a dashed curve.
Comparing it with the numerically obtained stationary solution of
Eq.~(\ref{GPn}) with $g = -(2 \pi)^{3/2}$ (solid curve), we see that the
central density is larger and the tail is longer for the stationary
solution than for the Gaussian function.

Since $R$ cannot directly be measured in experiments, we rewrite the
equation of motion in terms of the central column density
\begin{equation}
D = 2 \int_0^\infty dr |\psi_{\rm G}|^2 = \frac{1}{\pi R^2}.
\end{equation}
We consider small deviations from $g_0 = -(2 \pi)^{3/2}$ and $R_0 = 1$,
and decompose $D$ and $g$ as $D = D_0 + \tilde D$ and $g = g_0 + \tilde
g$, where $D_0 = 1 / \pi$.
Equation (\ref{Rddot}) can then be rewritten as
\begin{equation} \label{ddotd}
\ddot{\tilde D} = \tilde D - \frac{1}{\sqrt{2} \pi^{5/2}} \tilde g,
\end{equation}
where we have kept only the terms linear in $|\tilde D|$ and $|\tilde
g|$.
From Eq.~(\ref{feedback}), we have
\begin{equation} \label{dotg}
\dot{\tilde g} = A \tilde D + B \dot{\tilde D} + C \ddot{\tilde D}.
\end{equation}
Differentiating Eq.~(\ref{ddotd}) with respect to $t$ and using
Eq.~(\ref{dotg}), we obtain
\begin{equation} \label{mat}
\frac{d}{dt} \left( \begin{array}{c} \tilde D \\ \dot{\tilde D} \\
\ddot{\tilde D} \end{array} \right) = \left( \begin{array}{ccc} 0 & 1 & 0
\\ 0 & 0 & 1 \\ -\frac{A}{\sqrt{2} \pi^{5/2}} & 1 - \frac{B}{\sqrt{2}
\pi^{5/2}} & -\frac{C}{\sqrt{2} \pi^{5/2}} \end{array} \right)
\left( \begin{array}{c} \tilde D \\ \dot{\tilde D} \\
\ddot{\tilde D} \end{array} \right).
\end{equation}

If the $3 \times 3$ matrix in Eq.~(\ref{mat}) has an eigenvalue whose real
part is positive, $\tilde D$ diverges exponentially in time.
Therefore, the condition for stability is that all the eigenvalues have
negative real parts, which can be examined by using the Routh--Hurwitz
criterion~\cite{Routh} without actually solving the eigenvalue equation.
The stability condition of Eq.~(\ref{mat}) is found to be (see
Appendix~\ref{app} for derivation)
\begin{subequations}
\label{stabcond}
\begin{eqnarray}
A & > & 0, \\
B & > & \sqrt{2} \pi^{5/2}, \\
C & > & 0, \\
A & < & C \left( \frac{B}{\sqrt{2} \pi^{5/2}} - 1 \right).
\end{eqnarray}
\end{subequations}
These inequalities qualitatively agree with the stability diagram in
Fig.~\ref{f:stability}.

\section{Effects of experimental imperfections on the stability of
feedback control}
\label{s:noise}

We have so far assumed that the peak column density $D$ can be measured
precisely.
However, in real experiments, there are always imperfections in the
measurement, such as errors and the finite resolution of time and space.
In this section, we investigate the effects of such imperfections on the
stability of our feedback control.

\begin{figure}[tb]
\includegraphics[width=8.4cm]{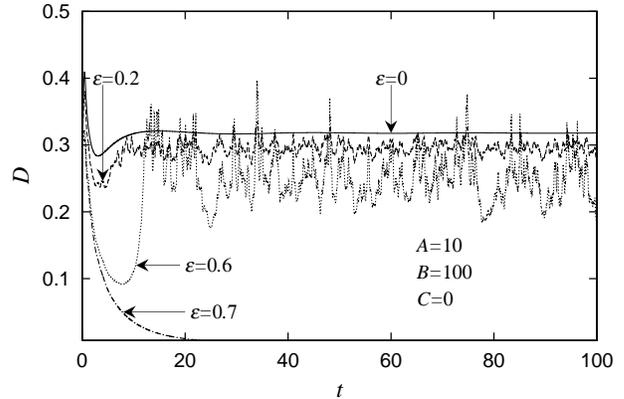}
\caption{
Time evolution of the peak column density $D$ with measurement errors
given in Eq.~(\ref{Derr}) for $\varepsilon = 0$ (solid line), $\varepsilon
= 0.2$ (dashed line), $\varepsilon = 0.6$ (dotted line), and $\varepsilon
= 0.7$ (dot-dashed line).
The feedback parameters are $A = 10$, $B = 100$, and $C = 0$.
The other conditions are the same as in Fig.~\ref{f:evolution}.
}
\label{f:noise}
\end{figure}

We first study the effects of measurement errors of $D$ on the stability.
We assume that the measured value with an error is given by
\begin{equation} \label{Derr}
D_\varepsilon = D (1 + \varepsilon v_{\rm rnd}),
\end{equation}
where $D$ is defined by Eq.~(\ref{D}), $\varepsilon$ describes the error
level, and $v_{\rm rnd}$ is a random variable that simulates the
measurement error and is assumed to obey the normal distribution
$e^{-v_{\rm rnd}^2 / 2} / \sqrt{2 \pi}$.
Figure~\ref{f:noise} shows some examples of the time evolution of $D$, in
which $D_\varepsilon$ is used in the feedback equation (\ref{feedback})
instead of $D$.
We find that the system is tolerant against an error level up to about
60\% in every measurement, and the stability is excellent.
In Fig.~\ref{f:noise}, we see that $D$ has a tendency to decrease with an
increase in $\varepsilon$.
This phenomenon is similar to that in the case of oscillating
interactions~\cite{SaitoL,Ab03}, where the peak density is suppressed
by the oscillation of the interaction.
In the present case, the random fluctuations in $g$ play the role of
oscillating interactions.

\begin{figure}[tb]
\includegraphics[width=8.4cm]{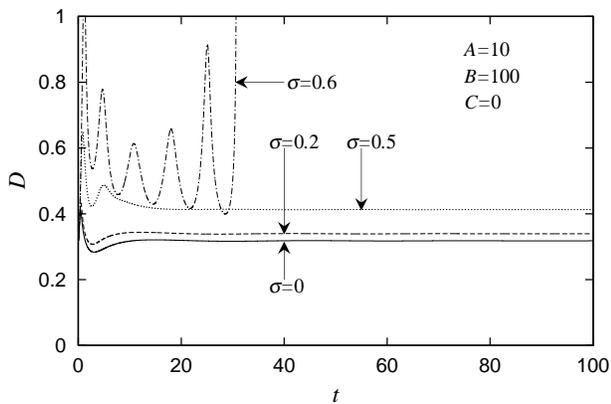}
\caption{
Time evolution of the peak column density $D$ with spatial resolution
given in Eq.~(\ref{Dsigma}) for $\sigma = 0$ (solid line), $\sigma = 0.2$
(dashed line), $\sigma = 0.5$ (dotted line), and $\sigma = 0.6$
(dot-dashed line).
The feedback parameters are $A = 10$, $B = 100$, and $C = 0$.
The other conditions are the same as in Fig.~\ref{f:evolution}.
}
\label{f:res}
\end{figure}

We next study the effect on the stability of the spatial resolution in the
measurement of $D$.
We assume that due to the finite spatial resolution, the measured value is
filtered by a Gaussian function,
\begin{equation} \label{Dsigma}
D_\sigma = \int dx dy \frac{1}{2 \pi \sigma^2} e^{-\frac{x^2 + y^2}{2
\sigma^2}} d_{\rm col}(x, y),
\end{equation}
where $d_{\rm col}(x, y)$ is given in Eq.~(\ref{dc}) and $\sigma$
characterizes the spatial resolution.
Figure~\ref{f:res} shows the time evolution of $D$, in which $D_\sigma$ is
used in Eq.~(\ref{feedback}) instead of $D$.
It is remarkable that the stability is very robust against low
resolution.
In fact, Fig.~\ref{f:res} shows that the acceptable resolution can be
almost the size of the condensate itself.

We have assumed so far that the successive measurements are performed at
an interval of $\Delta t = 0.01$.
We have also examined the stability for larger $\Delta t$ with $A = 10$,
$B = 100$, and $C = 0$ (with the other conditions the same as in
Fig.~\ref{f:evolution}), and found that stability is achieved for a time
interval of up to $\Delta t = 0.4$.

As an example, let us consider the case of $^{85}{\rm Rb}$ atoms and take
the units of length and time to be $3.5 \mu{\rm m}$ and $16$ ms, which
corresponds to $\omega = 10 \times 2 \pi$ Hz.
Then, the resolution of the phase-contrast imaging must be $\sigma
\lesssim 1.7 \mu{\rm m}$, and the interval of the measurements must be
$\Delta t \lesssim 6.4$ ms, which corresponds to a frame rate $\gtrsim
160$ Hz.
If we use a larger condensate (i.e., a smaller $\omega$), these
restrictions can be relaxed.

\section{Conclusions}
\label{s:conc}

We have shown that a BEC droplet (self-trapped condensate) can be
stabilized in 3D free space by the feedback control of the strength of the
interaction between atoms.
By negative feedback on the strength of the interaction through
nondestructive monitoring of the peak column density of the condensate, we
can prevent the condensate from collapsing and expanding.
Even starting from a Gaussian wave function, we can reach the stationary
state of the GP equation by feedback control.

We have considered the feedback from the peak column density $D$ and
its time derivatives $\dot{D}$ and $\ddot{D}$ (Eq.~(\ref{feedback})),
and have examined the stability of the system for various values of the
parameters.
We found that stability is obtained for a wide range of parameters, as
shown in Figs.~\ref{f:depend} and \ref{f:stability}.

We have also investigated the stability against experimental
imperfections, such as measurement errors (Fig.~\ref{f:noise}), finite
spatial resolution (Fig.~\ref{f:res}), and finite time intervals between
measurements.
We have found that the stability is robust against these imperfections.

In this paper, we have considered only the simplest form of negative
feedback.
More robust stability may be obtained and the stationary state may be
reached more quickly if more sophisticated methods are used, such as the
Kalman filter~\cite{Kalman} and robust control~\cite{robust}.

\begin{acknowledgments}
This work was supported by Grant-in-Aids for Scientific Research (Grant
No.\ 17740263, No.\ 17071005, and No.\ 15340129) and by the 21st Century
COE programs on ``Coherent Optical Science'' and ``Nanometer-Scale
Quantum Physics'' from the Ministry of Education, Culture, Sports, Science
and Technology of Japan.
M.U. acknowledges support by a CREST program of the JST.
\end{acknowledgments}

\appendix

\section{Stability condition with the Routh--Hurwitz criterion}
\label{app}

According to the Routh--Hurwitz criterion~\cite{Routh}, all solutions of a
polynomial equation
\begin{equation}
a_0 \lambda^n + a_1 \lambda^{n-1} + \cdots + a_{n-1} \lambda + a_n = 0
\end{equation}
have negative real parts if (1) all the coefficients $a_i$ for $i = 0, 1,
\cdots, n$ are real and positive and (2) the determinants
\begin{equation}
\left| \begin{array}{ccccc}
a_1 & a_3 & a_5 & \cdots & a_{2i-1} \\
a_0 & a_2 & a_4 & \cdots & a_{2i-2} \\
0 & a_1 & a_3 & \cdots & a_{2i-3} \\
0 & a_0 & a_2 & \cdots & a_{2i-4} \\
\vdots & \vdots & \vdots & \ddots & \vdots \\
0 & \cdots & \cdots & \cdots & a_i
\end{array} \right|
\end{equation}
for $i = 2, 3, \cdots, n$ are positive.

The system described by Eq.~(\ref{mat}) is stable if all the eigenvalues
of the $3 \times 3$ matrix on the right-hand side have negative real
parts.
The eigenvalue equation is given by
\begin{equation} \label{evaleq}
\lambda^3 + \frac{C}{\sqrt{2} \pi^{5/2}} \lambda^2 +
\left(\frac{B}{\sqrt{2} \pi^{5/2}} - 1 \right) \lambda + \frac{A}{\sqrt{2}
\pi^{5/2}} = 0.
\end{equation}
Applying the Routh--Hurwitz criterion to Eq.~(\ref{evaleq}), we obtain the
condition for the stability given in Eq.~(\ref{stabcond}).

\end{document}